\documentclass[usenatbib,usegraphicx,letters]{mn2e}

\usepackage{amssymb}
\usepackage{txfonts}
\newcommand{\aap}{A\&A}
\newcommand{\aaps}{A\&AS}

\newcommand{\aj}{AJ}
\newcommand{\apj}{ApJ}

\newcommand{\apjs}{ApJS}
\newcommand{\mnras}{MNRAS}
\newcommand{\pasp}{PASP}

\newcommand{\ao}{AO\,0235$+$164}%
\def\phn{\phantom{0}}

\begin{document}
\voffset=-0.8in
\title[Photo-polarimetry of \ao]{Extreme photo-polarimetric behaviour of the
  blazar \ao%
\thanks{Based on observations collected at the Centro
  Astron\'omico Hispano Alem\'an (CAHA), operated jointly by the Max-Planck
  Instituf f\"ur Astronomie and the Instituto de Astrof\'{\i}sica de
  Andaluc\'{\i}a (CSIC)}
}

\author[Cellone et al.]{Sergio
  A. Cellone$^{1,2}$\thanks{scellone@fcaglp.unlp.edu.ar}, Gustavo
  E. Romero$^{1,3}$, Jorge A. Combi$^{3,4}$ and Josep Mart\'{\i}\,$^{4}$\\
$^1$Facultad de Ciencias Astron\'omicas y Geof\'\i sicas, Universidad Nacional
de La Plata, Paseo del Bosque, B1900FWA La Plata, Argentina\\
$^2$IALP, CONICET--UNLP, Argentina\\
$^3$Inst.\ Argentino de Radioastronom\'{\i}a, C.C.5, 1894 Villa
Elisa,  Buenos Aires, Argentina\\
$^4$Universidad de Ja\'en, Campus Las Lagunillas s/n Ed-A3, 23071-Ja\'en,
  Spain
}

\maketitle

\begin{abstract}
We present optical photo-polarimetric observations with high temporal
resolution of the blazar \ao. Our data, the first to test the
photo-polarimetric behaviour of this object at very short time-scales,
show significant micro-variability in total flux, colour index, linear
polarization degree, and position angle. Strong inter-night variations are
also detected for these parameters.
Although no correlation between colour index and total flux was found, our
data seem to support the general bluer-when-brighter trend already known for
this object.
The polarization degree, in turn, shows no correlation with total flux, but
a clear trend in the sense that colour index is redder (the spectrum is
softer) when the measured polarization is higher.

\end{abstract}


\section{Introduction}
 \label{s_intro}

The BL\,Lacertae object \ao\ is one of the most intensively monitored and
variable blazars \citep[e.g.][]{WSL88}. Very rapid changes in its flux
density have been reported across the entire electromagnetic spectrum, from
radio frequencies \citep{QWK92,RCB97,KQL99} to gamma-ray energies
\citep{HBB99}.

 At optical wavelengths, rapid variations of a few tenths of a magnitude
within a single night have been repeatedly detected \citep{RMSS96, HW96,
NM96}. \citet{RCC00aa} found the most extreme variability event ever
reported for this object: changes up to 0.5 mag in the $R$ and $V$ bands were
detected within a single night, whereas night-to-night variations reached up
to 1.2 mag.

 The historical optical light-curve of \ao\ has been compiled by
\citet{FL00}, while several international follow-up campaigns have provided
detailed multiwavelength monitoring for this object \citep{RVA01,
RVI05}. Those data suggested the existence of a $\sim 5.7$\,yr
quasi-periodical behaviour, then leading to an interpretation in terms of a
binary supermassive black hole system at the core of the source
\citep{RFN03, OVR04}. Recent multifrequency observations by \citet{RVK06},
however, have suggested a longer ($\sim 8$\,yr) periodicity.

On the other hand, since the optical emission in blazars is expected to be
dominated by synchrotron radiation from the relativistic jet, optical
polarimetry is a useful tool to probe the innermost regions of these
objects, especially when high time-resolution data are obtained
\citep{ACR03}. In spite of this fact, polarimetric observations of blazars
are still scarce, and the first attempts to characterise the polarimetric
micro-variability of different blazar classes have just recently been done
\citep{ARC05}. In the particular case of \ao, previous studies revealed high
degrees of optical polarization \citep[e.g.,][]{MBB90}, with inter-night
random fluctuations \citep{TSN92}, but nothing was known about its
polarimetric behaviour on very short time-scales.

In this Letter we present the first insight into the polarimetric
micro-variability of \ao. Our high-temporal resolution data show significant
variations both in the linear polarization degree and the position angle at
intra-night as well as at inter-night time-scales. Simultaneous differential
photometry (at the $B$ and $R$ bands) shows that no apparent correlation
exists between variations in polarized and total flux. However, there is a
clear trend in the sense that the ($B-R$) colour gets redder (i.e., the
spectrum gets softer) when the polarization degree is higher.

We present our observational data in Section~\ref{s_obsred}, we give the
photometric and polarimetric variability results in Section~\ref{s_resu},
and we close with a short discussion and our conclusions in
Section~\ref{s_discu}.

\section{Observations and data reduction}
 \label{s_obsred}

We observed \ao\ along 6 nights in November 2005 and two additional nights
in December 2005, using the Calar Alto Faint Object Spectrograph (CAFOS) in
its imaging polarimetry mode, at the Calar Alto (Spain) 2.2\,m telescope.
In this instrumental setup, a polarizer unit, consisting of a Wollaston
prism %
plus a rotatable half-wave plate, is inserted into the light beam, thus
producing two orthogonally polarized images of each object on the focal
plane, separated by 18.9 arcsec.
The detector was a SITE\#1d CCD, with $2\mathrm{k} \times 2\mathrm{k}\;
24\,\mu$m pixels, a gain of 2.3 electrons adu$^{-1}$, and a read-out noise of
5.06 electrons.

In order to avoid any possible overlapping of the ordinary (O) image of one
object with the extraordinary (E) image of a different object, a mask with
alternate blind and clear stripes is placed before the detector.  Hence,
each polarimetric frame consists of alternate O and E $\sim 20$ arcsec wide
stripes.  This procedure also enhances the S/N ratio by reducing the sky
contribution to half its otherwise value, although, as a drawback, half of
the field of view is lost.

Four frames, each taken with a different position of the  half-wave
plate ($0^\circ$, $22.5^\circ$, $45^\circ$, $67.5^\circ$) are needed to
obtain the normalised Stokes parameters ($U$, $Q$) for linear
polarization. The relevant expressions for $U$ and $Q$ can be found in,
e.g., \citet{LH99e} and \citet{ZCB05}. From them, the degree of linear
polarization and its corresponding position angle are calculated in the
usual way:
\begin{equation}
P = \sqrt{Q^2 + U^2} ,
\label{e_pol}
\end{equation}
\begin{equation}
\Theta = \frac{1}{2} \, \arctan \left( {U \over Q} \right).
\label{e_theta}
\end{equation}

On the other hand, by adding the fluxes corresponding to the O and E images
of the same object within each frame, photometric data are also
obtained. This allowed us to simultaneously study both the photometric and
the polarimetric behaviours of our target with time. Hence, our
observational program consisted of unit blocks with the following form:
\begin{itemize}
\item Four consecutive frames, with the polarizer unit in, with half-wave
  plate angles $0^\circ$, $22.5^\circ$, $45^\circ$, and $67.5^\circ$,
  respectively, all obtained trough an $R$ (Cousins) filter. From these we
  obtained one polarimetric data point and 4 photometric, $R$-band, data
  points.
\item One $B$ frame, without the polarizer unit. This gave us a photometric
$B$-band data point.
\end{itemize}

A sequence of $N$ such 5 frames blocks was obtained for \ao\ each night,
with $3 \le N \le 11$, depending on the observing conditions. This procedure
gave us $N$ polarimetric ($R$-band) data points, $N$ photometric $B$ data
points, and $4N$ photometric $R$ data points per night. The exception is the
night of Nov/03--04, 2005, when just two photometric $R$ data could be
obtained, due to bad weather.  Individual integration times ranged from 360
to 450\,s for $B$, and between 240 and 360\,s for $R$ (all 4 $R$-band frames
within a given polarimetric block were obtained with identical exposure
times).
Several standard highly-polarized and unpolarized stars from \citet{TBW90}
were also observed (at least two polarimetric cycles each) for calibration
purposes.

Atmospheric conditions were dissimilar along the whole observing time,
ranging from high-quality photometric nights to rather mediocre conditions
due to poor seeing and/or passing cirrus. However, given that both polarimetry
and differential photometry are robust against poor observing conditions,
these had no systematic impact on our data, except for a loose in
S/N ratio. We shall further discuss this below (see Section~\ref{s_resu}).

\begin{figure}
\centering
\includegraphics[width=0.85\hsize]{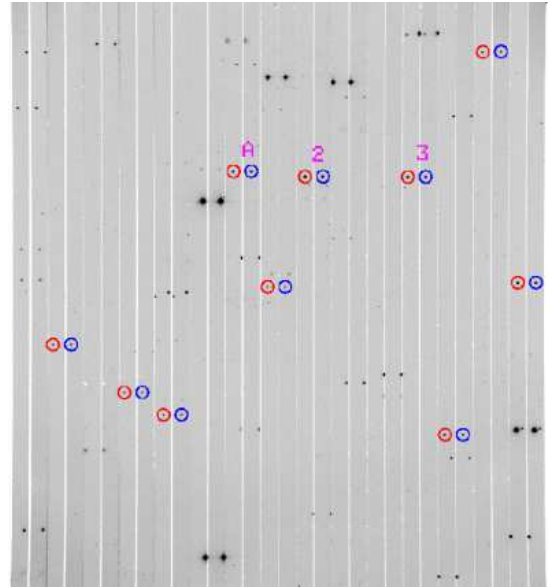} %
\caption{Polarimetric frame of the observed field ($R$-band). For each
measured object, red and blue circles mark the ordinary and extraordinary
images, respectively. \ao\ (A) as well as stars used for differential
photometry are labelled. East is up, North is to the right.}
\label{f_field}
\end{figure}

All science frames were bias-corrected using a master bias prepared from 30
individual bias frames, and flat-fielded using twilight sky flats. Standard
routines within the \textsc{iraf}\footnote{%
IRAF is distributed by the National Optical Astronomy Observatories,
    which are operated by the Association of Universities for Research
    in Astronomy, Inc., under cooperative agreement with the NSF, USA.}%
 package were used for image processing and
all subsequent data extraction.

We obtained the instrumental magnitudes corresponding to both the O and E
images of \ao\ on each frame, using the aperture photometry task
\textsc{apphot}. The same was done for 9 stars evenly distributed on the
field and suitably placed within the mask stripes (i.e., we rejected stars
close to the edge of a stripe). These stars were used to evaluate the
instrumental/foreground polarization, while two of them were used to
construct the differential photometry light-curves (see
Section~\ref{s_resu}).
All the measured objects in the field are shown in Fig.~\ref{f_field}; for
each object, red and blue circles mark the O and E images, respectively.
Our target (\ao) as well as the two stars used for differential photometry
(see Section~\ref{ss_phot}) are labelled.

A 3 arcsec radius aperture was always used.  Blazars photo-polarimetry can
be affected by spurious variations due to the host galaxy \citep{CRC00,
A06}, but this should not be relevant for \ao\, given its relatively high
redshift ($z=0.94$). However, a possible error source is an AGN lying $\sim
2$ arcsec south of \ao\ \citep[named ELISA in][]{RVI05}, which we resolve on
our best-seeing images, but appears merged with our target on the rest. We
checked for any spurious effect on the photo-polarimetric variability of
\ao\ by comparing data obtained through different apertures against each
other. Except for a change in S/N ratio, there was no significant
difference. We also checked for any correlation between photo-polarimetric
variability and changes in the seeing full-width at half-maximum (FWHM),
finding none.  Hence, we conclude that no systematic errors due to this
nearby object significantly affect our variability results.  However,
constant shifts are expected in total and polarized flux due to ELISA
and a faint absorbing galaxy $\sim 1\farcs3$ east from \ao\ \citep[G1
in][]{NCP96}, respectively. We shall return to this point in
Section~\ref{s_resu}.

A local sky value was measured, as usual, within an annulus surrounding each
aperture. Since both O and E images of the same source are close
($\sim20$\,arcsec) to each other on the polarimetric frames, all pixels
corresponding to the E image (including sky) were masked out when measuring
the O image, and vice-versa.

\section{Results}
 \label{s_resu}

\subsection{Photometry}
 \label{ss_phot}

The differential light-curve for \ao\ was obtained in the usual way, using a
non-variable star (here named star~2) in the field as comparison
\citep[e.g.,][]{HJ86}, while another star (3) was used to construct a second
differential light-curve against star~2, to be used for control purposes
(see Fig.~\ref{f_field}).  These two stars are the same ones we used in our
previous studies of \ao\ \citep{RCC00aa, RCCA02}; star~2 is also star~8 in
\citet{SBH85} and star~10 in \citet{GKM01}.

\begin{figure}
\includegraphics[width=0.9\hsize]{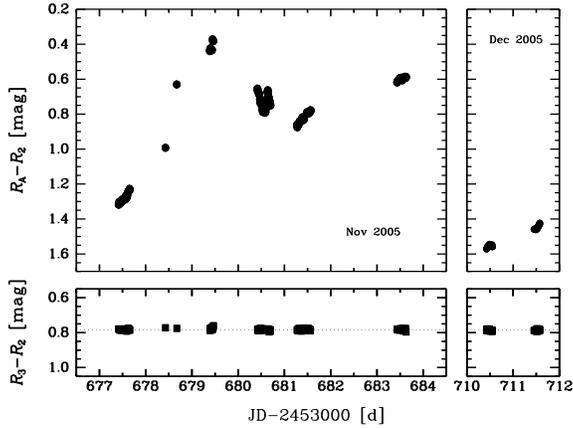} %
\caption{$R$-band differential light-curve for \ao\ vs.\ comparison star
  (upper panels) and for control-star vs.\ comparison star (lower
  panels) for the whole campaign (left: Nov.\ 2005, right: Dec.\ 2005).}
\label{f_photoR}
\end{figure}

\begin{figure}
\includegraphics[width=0.9\hsize]{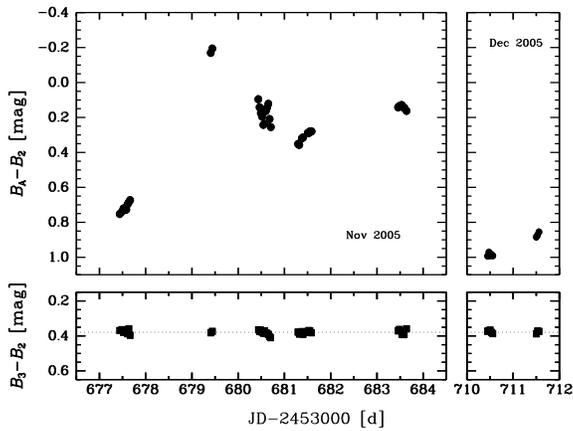} %
\caption{Same as Fig.~\ref{f_photoB} for the $B$-band.}
\label{f_photoB}
\end{figure}

The results are shown in Figures~\ref{f_photoR} and \ref{f_photoB} for the
$R$ and $B$ bands, respectively. Strong inter-night as well as significant
intra-night variations are clearly seen, with similar behaviours in both
bands. Note the stability of the control light-curves (lower panels). The
statistical significances of intra-night variations were assessed following
\citet{HWM88}, i.e.\ defining a scaled confidence parameter $C \Gamma^{-1}$
which depends on the dispersions of both light-curves (target--comparison
and control--comparison) and on a corrective factor that equalises their
respective instrumental errors \citep[see also][]{CRA07}.
Table~\ref{t_fotodif} gives the date (column 1), number of $R$-band data
points (column 2), target--comparison light-curve dispersion (column 3),
control--comparison light-curve dispersion (column 4), scaled confidence
parameter (column 5), and variability classification, which is
``\textsc{yes}'' when $C \Gamma^{-1} \ge 2.576$ (column 6).

\begin{table}
\caption{Variability parameters for the $R$ light-curve on different nights.}
\label{t_fotodif}
\begin{tabular}{lccccc}
\hline
\hline
Night & $N$ & $\sigma_{A-2}$ & $\sigma_{3-2}$ & $C \Gamma^{-1}$ & Variable? \\
      &     &  mag           &  mag           & & \\
\hline
Nov/02-03  &    \phn32 & 0.026  &  0.004  & \phn\phn6.559 & \textsc{yes} \\
Nov/03-04  & \phn\phn2 & 0.256  &  0.003  &       102.992 & \textsc{yes} \\
Nov/04-05  &    \phn12 & 0.026  &  0.007  & \phn\phn3.980 & \textsc{yes} \\
Nov/05-06  &    \phn44 & 0.042  &  0.004  &    \phn11.065 & \textsc{yes} \\
Nov/06-07  &    \phn40 & 0.030  &  0.004  & \phn\phn7.698 & \textsc{yes} \\
Nov/08-09  &    \phn28 & 0.008  &  0.004  & \phn\phn1.942 & \textsc{no} \\
Dec/05-06  &    \phn20 & 0.008  &  0.005  & \phn\phn1.570 & \textsc{no} \\
Dec/06-07  &    \phn16 & 0.013  &  0.005  & \phn\phn2.556 & \textsc{no} \\[2pt]
Whole campaign & 194 & 0.357  &  0.005  & \phn78.327  & \textsc{yes} \\
\hline
\end{tabular}
\medskip
\end{table}

For the $1^\mathrm{st}$, $4^\mathrm{th}$, and $5^\mathrm{th}$ nights we have
well-sampled light-curves displaying significant intra-night variability.
We get a formally very high value of $C \Gamma^{-1}$ for the $2^\mathrm{nd}$
night, although with only two data points in $R$. Despite this poor sampling
due to bad weather, these data follow the inter-night trend: \ao\ brightened
by $\sim 0.95$\,mag from the $1^\mathrm{st}$ to the $3^\mathrm{rd}$
nights. During the last night in November and both December nights, \ao\
displayed no significant intra-night variability.

We also calculated $R$ and $B$ magnitudes for \ao\ in the standard
system. To do so, we used photometry of field stars from our own previous
data \citep[star 2: $R_2=15.828$,][]{RCC00aa} and from the literature
\citep{SBH85, GKM01}.  During our observations, \ao\ ranged between $16.2
\lesssim R \lesssim 17.4$ mag; i.e., it was at a state of moderate
brightness as compared to its historical light-curve, and about 2 mag fainter
than during its major outbursts \citep[e.g.,][]{RVI05}.
$B-R$ colour-indices were calculated by interpolating $R$ and $B$ magnitudes
to common time instants.

We verified that subtracting the flux contribution from ELISA and correcting
for Galactic plus foreground absortion does not change our micro-variability
results in any significant way, except for nearly constant shifts in the
light curves ($\sim 1.2$ and $\sim 1.8$\,mag brighter in $R$ and $B$,
respectively). However, colour index is significantly affected. We thus
obtained a corrected colour index $(B-R)_0$ following \citet{RVI05}, and we
transformed it to spectral index $\alpha$ ($F_\nu \propto \nu^{-\alpha}$),
using $\lambda_B=0.44\, \micron$, $\lambda_R=0.64\, \micron$ \citep{B79}.

\subsection{Polarimetry}
 \label{ss_pol}

We obtained the linear polarization percentage ($P$) and position angle
($\Theta$) for \ao\ using equations~(\ref{e_pol}) and (\ref{e_theta}), as
outlined in Section~\ref{s_obsred}. The contribution from instrumental
polarization was derived from data on unpolarized standard stars, while
foreground polarization was estimated from star 2, which lies at
$76$\,arcsec from \ao. Foreground polarization is small ($<0.5\,\%$), in
agreement with the low Galactic extinction towards \ao. We also checked for
any dependence of instrumental polarization with position on the CCD. Stars
near the border of the field tend to display relatively large $P$ values;
however, since \ao\ was always centrally placed on the field, no systematic
effect is expected on our polarimetric curves.  The position angle was
transformed to the Equatorial system using data from highly polarized
standard stars.

\begin{figure}
\includegraphics[width=0.95\hsize]{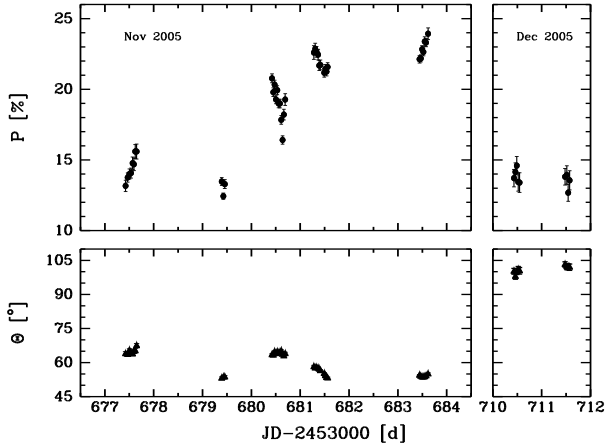} %
\caption{$R$-band polarization (upper panels) and position angle (lower
  panels) against time for the whole campaign.}
\label{f_poltheta}
\end{figure}

\begin{figure}
\includegraphics[width=0.95\hsize]{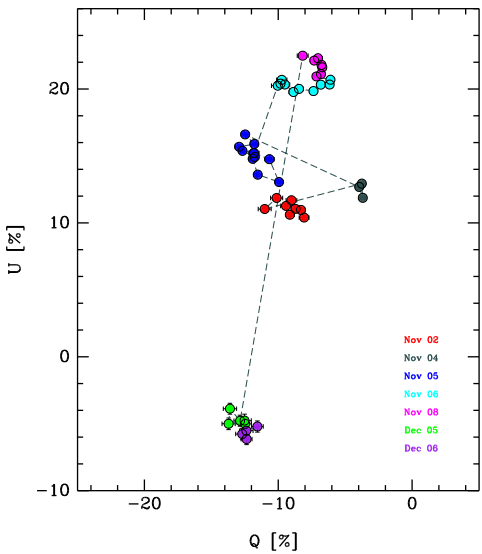} %
\caption{Evolution of the polarization on the Stokes plane.
Different colours are used for data corresponding to different nights.}
\label{f_qu}
\end{figure}

Fig.~\ref{f_poltheta} shows the behaviour of $P$ and $\Theta$ along the
whole campaign. The degree of polarization was high ($12.5\,\% \lesssim P
\lesssim 24.0\,\%$) and clearly variable, both at intra- and inter-night
time-scales. The position angle was also variable, although no clear
connection between the behaviours of both parameters can be seen. In fact,
some of the nights \ao\ displayed conspicuous variability in $P$ without any
significant change in $\Theta$, while the converse is also true.

The significance of $P$ and $\Theta$ variability for \ao\ can be assessed by
comparing it with the behaviour of these same parameters for field stars. We
found that the dispersions of the $P$ and $\Theta$ curves against time for
the stars strongly depend on their magnitudes, as expected since in this
case any variability should be due to errors. From these relations, we
estimated the dispersions in polarization degree and position angle for \ao,
expected just from errors, to be $0.10\,\% \lesssim \sigma_P \lesssim
0.39\,\%$ and $0.23^\circ \lesssim \sigma_\Theta \lesssim 0.84^\circ$,
respectively.
As a result, 5 out of 7 nights with polarimetric data available show
statistically significant ($>3\sigma$) microvariability in $P$, while just
two of them show significant microvariability in $\Theta$.
Note that intra-night variability amplitudes up to $\Delta P \simeq 4.4\,\%$
($\sim 8\, \sigma_P$) and $\Delta \Theta \simeq 5^\circ$ ($\sim 6\,
\sigma_\Theta$) were detected. Dust absortion from the foreground galaxy G1
would introduce a constant shift in $P$ amounting to, at the very most, $\sim
2.8\,\%$, but with no variability effect.

\begin{figure}
\includegraphics[width=0.95\hsize]{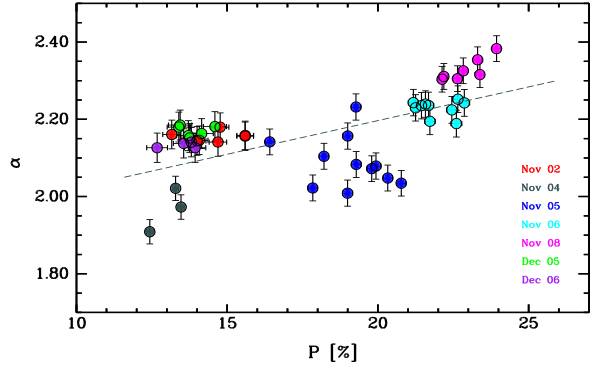} %
\caption{Spectral index against polarization.
Different colours are used for data corresponding to different nights. The
   total least-squares fit (Eq.~\ref{e_Palpha}) is shown with a dashed line.}
\label{f_polcol}
\end{figure}

The behaviour of \ao\ on the Stokes plane is shown in Fig.~\ref{f_qu}. Note
the different location of the Dec/2005 data ($Q \approx -13\,\%$, $U \approx
-5\,\%$) as compared to those at the beginning of our campaign, despite
their similar $P$ values. This underscores that there is no correlation
between $P$ and $\Theta$. Neither there is any correlation between $P$ and
the source brightness or the variability gradient of the observed flux.

A well-defined correlation appears to exist, instead, between $P$ and the
corrected colour index $(B-R)_0$ (or spectral index $\alpha$). As shown in
Fig.~\ref{f_polcol}, $P$ tended to be lower when the spectrum was harder,
although the relation displays a significant scatter.

\section{Discussion and conclusions}
 \label{s_discu}

\citet{FL00} suggest a two-component origin for the optical emission of \ao,
given that its colour index vs.\ magnitude relation seems to show two
different behaviours, depending on the overall flux level of the source.
Previous micro-variability data \citep{RCC00aa} have suggested, although
with a low confidence level, a mean trend in the sense that the optical
emission of \ao\ became brighter when its spectrum was harder.  Our present
data span rather limited ranges both in magnitude and colour index [$\Delta
R \simeq 1.4$\,mag; $\Delta(B-R)_0 \simeq 0.19$\,mag], so this may be the
reason why we did not detect any definite trend between colour index and
magnitude. In any case, our data fall at the low-luminosity end of the
relation shown in \citet[their fig.~7]{RVA01}, which displays a
redder-when-fainter behaviour. The relatively large scatter of this relation
would thus be intrinsic, since micro-variations usually do not follow the
general trend.

As mentioned in the previous section, we did not detect any clear
correlation between polarization degree and total flux. This lack of
correlation seems to be a common feature of microvariability in blazars
\citep[e.g.,][]{TDP01}. However, a well-defined relation between polarization
degree and spectral index does show up from our data
(Fig.~\ref{f_polcol}). Note that this relation is in the opposite sense as
expected if the decrease in polarization degree were due to a higher
relative contribution from the (redder, unpolarized) host galaxy light.

The spectral index ranged from
1.91 to 2.38, and we obtained the following linear relation against
polarization degree:
\begin{equation}
P (\%) = (56.3 \pm 1.4)\, \alpha - (103.7 \pm 3.1)\, .
\label{e_Palpha}
\end{equation}
This relation is in qualitatively good agreement with those found for
3C\,66A \citep{EP06} and OJ\,287 \citep{ESTS02}, suggesting that, despite
the different time-scales involved in those studies, a similar mechanism
could be operating in all three objects.  \citet{EP06} propose that the
polarization degree decreases when the ratio between the strength of the
regular and the chaotic components of the magnetic field decreases due to
blobs moving along the jet, thus breaking up the ordered structure of the
field. The higher degree of disorder in the magnetic field also leads to a
hardening of the spectrum.

Alternatively, \citet{BMBH90} suggested that the frequency dependency of the
polarization can be interpreted in terms of an inhomogeneous model, with two
components contributing to the optical emission. One is a polarized
component with a high-energy cut-off that could be identified with the
radiation produced by shock accelerated electrons in the compressed magnetic
field lines. The other component would have a steeper spectral index and
slight polarization and could correspond to the underlying jet
emission. Changes in the cut-off, due to variable energy losses, would lead
to a harder spectrum (lower $\alpha$) when the unpolarized component
dominates the optical emission.

\citet{RVK06} have recently presented observational support for an
additional component, in the form of a UV -- soft X-ray bump, lying between
the synchrotron and inverse Compton peaks in the spectral energy
distruibution of \ao. They suggest this extra component may arise from a
thermal accretion disk or from a distinct region in an inhomogeneous jet. In
either case, changes in $P$, spectral index and total flux are expected.

In the present Letter we have reported the existence of extremely rapid
variability in the optical flux and, for the first time, in the optical
polarization of the blazar \ao. There is no correlation between the
variability of both parameters, but a clear correlation is observed between
the spectral index and the polarization, in the sense that the polarization
is lower when the spectrum is harder. Such a behaviour might be the effect
of shocks moving through the inner jet of the source. These shocks could be
formed by collisions of relativistic plasma outflows with different
velocities. Alternatively, a two-component model, particularly if the UV --
soft X-ray bump has a thermal origin, could naturally lead to correlated
variability in polarization and spectral index.  Long-term polarization
observations could reveal whether this phenomenology is present on other
timescales, yielding then useful information
to further constrain the different models.

\section*{Acknowledgments}

We are deeply grateful to Ulli Thiele, Felipe Hoyo, Alberto Aguirre and the
whole Calar Alto staff for their invaluable help at the observatory.
We also thank the anonymous referee for helpful remarks.
This work received financial support from PICT\,13291 BID 1728/OC-AR
(ANPCyT) and PIP 5375 (CONICET), Argentina.
J.A.C. is a researcher of the programme {\em Ram\'on y Cajal} funded jointly
by the Spanish Ministerio de Educaci\'on y Ciencia (former Ministerio de
Ciencia y Tecnolog\'{\i}a) and Universidad de Ja\'en. We also acknowledge
support by DGI of the Spanish Ministerio de Educaci\'on y Ciencia under
grants AYA2004-07171-C02-02, FEDER funds and Plan Andaluz de Investigaci\'on
of Junta de Andaluc\'{\i}a as research group FQM322.


%
\end{document}